\documentclass[letterpaper,10pt]{article}
\newif\iftwocolumn
\usepackage[utf8]{inputenc}
\usepackage{hyperref}
 
\usepackage{makecell}
\usepackage{calc}
\usepackage[capitalise]{cleveref}
\usepackage{enumitem}
\usepackage[super]{nth}
\usepackage[mode=buildnew]{standalone}% requires -shell-escape
\usepackage{booktabs}
\usepackage{threeparttable}
\usepackage{multirow}
\usepackage{multicol}
\usepackage{pifont}% http://ctan.org/pkg/pifont
\usepackage{marvosym}
\usepackage{pdflscape}
\usepackage{afterpage}
\usepackage{wrapfig}
\usepackage{pgfplots}
    \usetikzlibrary{
        pgfplots.dateplot,
    }
\usepackage{xprintlen}
\pgfplotsset{compat=1.14}

\makeatletter
\newcommand\myparagraph{\@startsection{paragraph}{4}{\z@}%
                       {-8\p@ \@plus -4\p@ \@minus -4\p@}%
                       {-0.5em \@plus -0.22em \@minus -0.1em}%
                       {\normalfont\normalsize\bfseries\boldmath}}
\makeatother
\begin{document}
%-------------------------------------------------------------------------------

%don't want date printed
\date{October 2020}

\title{Cyber-Risks in Paper Voting}

%\title{\normalsize\Large \bf The Election-Fraudster's Cookbook:\\Cyber-Risk in Paper Voting}

%for single author (just remove % characters) USENIX FORMAT
% \author{}
\author{
{\rm David M.\ Sommer}\\
ETH Zurich
\and
{\rm Moritz Schneider}\\
ETH Zurich
\and
{\rm Jannik Gut}\\
ETH Zurich
\and 
{\rm Srdjan \v{C}apkun}\\
ETH Zurich 
}

% FC FORMAT

% \newif\ifshowauthors
% \showauthorsfalse
% \ifshowauthors
% \author{David M.\ Sommer
% \and Moritz Schneider
% \and Jannik Gut
% \and Srdjan \v{C}apkun
% \authorrunning{DM. Sommer, M. Schneider, J. Gut, S. \v{C}apkun}

% \institute{ETH Zurich}
% \else 
% \author{}
% \institute{}
% \fi

\maketitle

%-------------------------------------------------------------------------------
\begin{abstract}
Paper ballot voting with its fully-reviewable paper-trail is usually considered as more secure than their e-voting counterparts, given the large number of recent incidents.
% Considering a large number of vulnerabilities reported in the news and the importance of elections and referendums, the general public, as well as several security researchers, consider paper ballot voting with its fully verifiable paper trail as more secure than current e-voting alternatives. 
In this work, we explore the security of paper voting and show that paper voting, as it is implemented today, is surprisingly vulnerable to cyber-attacks. In particular, the aggregation methods of preliminary voting results of various countries rely on insecure communication channels like telephone, fax or non-secure e-mail. Furthermore, regulations typically do not mandate the use of secure channels for the transmission of preliminary results. 
We illustrate that preliminary results, despite their temporary nature, may have a severe impact on real-world decisions during the 3 to 30 days window until the final results are declared. An attacker exploiting this discrepancy can, e.g., benefit from stock market manipulation or call into question the legitimacy of the elections.
%We further introduce two new attacks: vote report delay and front-running, both of which can lead to different compromise of election results. 
This work investigates the cyber-risks in paper voting in a systematic manner by reviewing procedures in several countries (Estonia, France, Germany, the United Kingdom, and the United States of America) and through a comprehensive case-study of Switzerland. We examine the transmission systems currently in use through inquires from election officials. Moreover, we illustrate the feasibility of attacks by analyzing the frequent historical discrepancies between preliminary and final results. 
Considering our results and recent reports about easily modifiable preliminary results in Germany and the Netherlands, we conjecture similar weaknesses in other countries as well. 

% Their transmission prevalently over common digital communication channels like telephone, fax, or (non-secured) email, that do not provide message integrity or authenticity, exposes them to the risk of being intercepted, modified, or spoofed. 
%Furthermore, by providing several examples on the vast impact preliminary results have on financial and political decisions, e.g., the value of British Pounds after the Brexit referendum, we demonstrate possible attacks exploiting the time-gap between tampered preliminary results and the official paper-based reports.

\end{abstract}

%-------------------------------------------------------------------------------
\section{Introduction}
%-------------------------------------------------------------------------------
In most modern nations, the democratic process of electing representatives and deciding on critical matters directly is essential. Some countries only conduct elections, while others hold regular referendums as well.
%Historically, when implemented in stable democracies, ballot voting where the voters cast their ballots at voting stations is commonly perceived to be secure. 
The usually employed paper-based voting schemes, where voters cast their paper ballots at local voting stations, produce a verifiable paper-trail --- and are therefore commonly perceived as secure.
%Paper-based voting schemes --- where voters cast their paper ballots at local voting stations --- produce a verifiable paper-trail and are therefore commonly perceived as secure.
%Voters casting their paper ballots at local voting stations often perceive paper-based voting schemes as secure.
%The historically emerged confidence in paper-based voting schemes, where ... , 
%Ever since the emergence of paper-based voting schemes where voters cast their paper ballots at local voting stations, the confidence has raised to a level where it is commonly perceived as secure. 
The validity and integrity of paper voting has been studied extensively~\cite{aranha2016crowdsourced,martin2011,roth1998disenfranchised,voterregistration,stark2012evidence}. As the procedure from printing ballots, to collecting and counting, and finally, to the transmission of results is slow, cumbersome and prone to errors~\cite{mebane2004}, and with the increasing complexity of ballot propositions, interest has shifted to voting machines and internet-voting. However, the security of e-voting systems has been continuously challenged~\cite{bannet2004,kumar2011analysis,loeber2008voting}. Most effective attacks allowed modification of votes and compromise of the vote-counting process. The classic paper ballot voting, while being much slower, is therefore still seen as a reliable alternative and perceived as more trustworthy than its electronic counterparts.

Paper voting has been conducted for multiple decades, and robust systems have been established, mostly based on a decentralized vote casting and counting infrastructure. Usually, a local result is determined at a local voting office, followed by a paper report containing the exact counts that is hand-signed by the responsible official(s). These local results are then sent to the central office, where the final results are calculated. The communication of local results and/or ballots to the central office (typically via mail or dedicated transport), subsequent counting, and result validation can be slow and can take, depending on the country between 3 and 30 days. 
Due to this time-consuming process, most countries also allow preliminary results to be published, typically by transmitting local results using faster and unauthenticated communication channels (e.g., telephone, fax, email). 

Recent reports in Germany~\cite{pcwahl} and the Netherlands~\cite{dutchscandal} highlight negligent security engineering in widely used vote counting and aggregation software for preliminary results. We add to this discussion by systematically reviewing the paper voting process across several countries and investigating the problems that arise at the intersection of paper voting and digital transmission of its result. We show that, in several countries, the aggregation of preliminary results heavily depend on common computer technologies and is therefore exposed to a wide range of cyberattacks, possibly as much as e-voting. 

Interestingly, the regulations in the countries considered in this work do not require that authenticity or integrity of preliminary voting results is protected (e.g., \cite{germanyvotinglaw,estoniavotinglaw,francecodeelectoral,ukvotinglaw}). We examine these issues in Switzerland because it provides a vibrant democracy~\cite{swiss-abstimmungen} with around ten referendums per year, where the citizens decide directly about constitutional changes and controversial laws. In Switzerland, official preliminary results are published the same day the polls close, whereas the final results follow around 10 days later. Our study shows that the binary outcome of close national decisions can be flipped by changing the outcome of only a few municipalities, and that the normal error-fluctuations in vote-counting are sufficiently large to facilitate flipping-attacks.
% Considering the recent scandals about easily modifiable preliminary results in Germany~\cite{pcwahl} and the Netherlands~\cite{dutchscandal}, we conjecture similar weaknesses in other countries as well. 

In addition to naively tampering with the preliminary reports sent by voting stations, we introduce two new attacks: delaying reports and front-running real reports with fraudulent ones. The impact of these attack vectors may be smaller than report modification, but they are harder to mitigate.

%In contrast to paper voting, tampering attacks are already included in the adversary models of voting machines and internet-voting. However, 

It might seem that attacks on preliminary results are not so severe since these modifications will be corrected when the final results are published (i.e., when the correct results arrive via dedicated physical channels). We show, however, that this integrity time-gap between incorrect preliminary and correct final results can be damaging. It can be used for stock market manipulation, fraud, reduce the confidence in the government and democratic processes, or to delegitimize elected governments and referendum results. 
Most people are not aware of the official but much slower paper-trail and perceive the preliminary results as the final ones.

The immediate decay of the British Pound after the first interim result of the Brexit referendum~\cite{brexitlaw} or the risky German federal election in 2017 (see \Cref{sec:german_election}) highlights the severe impact that preliminary results have on trade markets and political decisions. 
%Our work, therefore, shows that the popular argument that only the final results are important to protect no longer holds in today's fast-paced world.
Therefore, our work shows the necessity to protect not only the final results but also the electronically transmitted preliminary results in today's fast-paced world. 

% \pagebreak
\myparagraph{Contributions.}
% Explain Paper voting process and Laws several jurisdictions
%  possible attacks
% one county in detail: Bescht land vo wält
%  * Tranmission by direct contact, several nknown to be weak
%  * Historic descrepancies of then years from public sources (newspaper statistics website), prelim deviate from final res
% impact on prelim results on real world., Brexit, Gold, DE politikum
% give recommendations based on our experience (WE are Go(o)d)
% 
% This work reviews typical paper voting mechanisms focusing on the aggregation of votes. After a general overview, we examine the legal obligations for preliminary vote accumulations in multiple jurisdictions (\cref{sec:understanding}). Next, \cref{sec:attack_vectors} introduces two new attack vectors, highlights recent vulnerabilities in actual transmission software, and illustrates the feasibility of attacks. 
% In the case study about Switzerland (\cref{sec:casestudy-switzerland}), we presents an exhaustive survey on the transmission systems in use, acquired from the responsible chancelleries directly. By comparing preliminary results published in news-articles to the official final results over a ten years period, we show frequent accumulation discrepancies that are sufficiently large to hide our proposed attacks.
% The real-world impact of fraudulent preliminary results is then discussed in \cref{sec:impact-preliminary-results}, and finally, we provide recommendations to improve the security of paper voting in \cref{sec:recommendations}.
\begin{itemize}[label=$\bullet$]
    \item This work reviews typical paper voting mechanisms, focusing on the aggregation of votes in several countries: Estonia, France, Germany, the United Kingdom, and the United States of America.
    \item We introduce two new attack vectors to tamper with preliminary results, front-running and delaying reports, and illustrate their feasibility. 
    \item In the case study about Switzerland, we present an exhaustive survey on the transmission systems in use, acquired from the responsible chancelleries by interview. We compare preliminary results published in news-articles to the official final results over a ten years period, and show frequent accumulation discrepancies that are sufficiently large to hide our proposed attacks.
    \item We show the real world impact of fraudulent preliminary results with, e.g., the pound exchange rate after Brexit.
\end{itemize}

\section{Understanding Paper Voting}
\label{sec:understanding}
% \begin{enumerate}
%     \item Paper voting
%     \begin{enumerate}
%         \item Voting
%         \item Counting
%         \item Transmission and Accumulation
%         \item final result
%     \end{enumerate}
%     \item indirect vs direct elections
%     \item Election vs referendum
%     \item Cyber in paper voting
%     \begin{enumerate}
%         \item pre-election polls / predictions \url{https://amp.theguardian.com/commentisfree/2019/may/20/mathematics-does-not-lie-why-polling-got-the-australian-election-wrong?__twitter_impression=true}
%         \item Fast updates wanted
%         \item Exit polls
%         \item preliminary results
%     \end{enumerate}
% \end{enumerate}

\begin{figure*}[tbp]
    \centering
    \iftwocolumn
        \resizebox{0.8\linewidth}{!}{\includestandalone{images/hierarchy}}
    \else
        \resizebox{\linewidth}{!}{\includegraphics{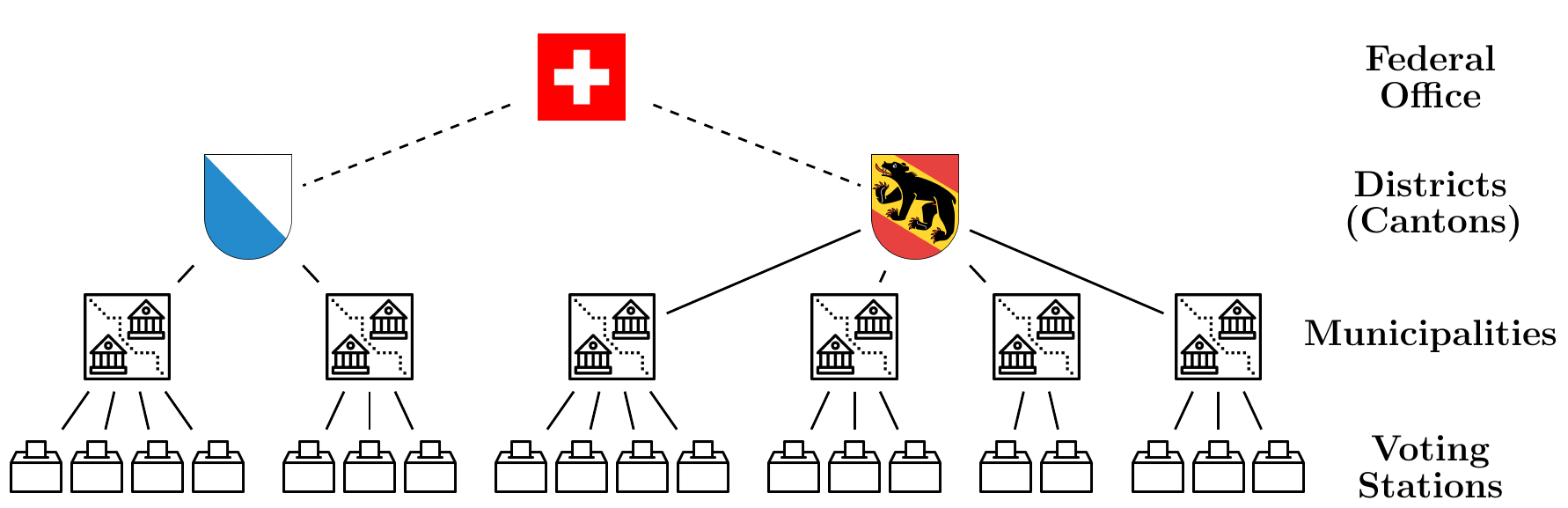}} 
    \fi
    \caption{Simplified vote aggregation infrastructure with Swiss hierarchical levels (districts are called 'cantons' in Switzerland). Similar accumulation structures are used in most countries. }
    % \spacesaver
    \label{fig:hirarchy}
    
\end{figure*}

Paper voting is the dominant voting mechanism used throughout most democratic countries. As its name indicates, paper voting relies on paper ballots, submitted by the voters either at the polling station or via mail. The counting and the transmission of results are then entirely performed on paper. The advantages are apparent: the results can be confirmed by verifying the physical paper trail. We review typical procedures based on laws and regulations from Estonia, France, Germany, the Netherlands, the United Kingdom, the United States of America, and Switzerland.

There are usually four stages of a vote that result in the final decision of the voters: the voting (ballot casting), counting, transmission, and accumulation. In the following, we will elaborate on these steps in a typical democracy.

\myparagraph{Voting.}
Voting in a paper-based voting scheme is usually conducted by filling out a paper ballot. Some ballots provide only simple options, e.g., yes and no, while others offer many choices and even write-ins to specify the name of the preferred candidate. The procedure of handing in ballots can widely differ between voting schemes. There are countries like France where any voter needs to physically go to a voting station, fill out the voting card, and drop it into a ballot box~\cite{francecodeelectoral}, while other countries like the United Kingdom~\cite{ukvotinglaw}, Germany~\cite{germanyvotinglaw}, and some states in the US additionally also allow postal voting. Recently some nations have started to use machine voting, where the vote is entered into a machine that produces a count in the end. Some variants of machine voting do not provide any paper trail, and therefore, do not allow for a manual recount. However, many voting machines produce a paper trail as well as a receipt to the voter to verify their vote was cast correctly. Nevertheless, these machines have been continuously proven to be insecure~\cite{kumar2011analysis,bannet2004,loeber2008voting}, allowing the attackers to modify the vote counts.
 
\myparagraph{Counting.}
In a paper ballot system, the votes must be manually counted by officials. Before opening the ballot boxes, unused ballots must usually be removed from the premise to reduce possible mistakes. Only then the ballot boxes are opened. Typically, the ballots are first ordered to the respective decision, e.g., into \emph{yes} and \emph{no} piles, and then counted manually (in some cases more than once). In the end, the counts are validated, e.g., by summing all votes and comparing it to the total number of ballots in the ballot box. Finally, all counts of the various ballot boxes of one voting station are accumulated, and a report with the results is compiled. The report, with all irregularities attached, is hand-signed by the supervising official(s). 

\myparagraph{Transmission.}
After the counting, the report must be transmitted to the superior instance as shown in Figure~\ref{fig:hirarchy}. Every office waits until it has received all results from its subordinates until it will issue another report summarizing all the results. This procedure continues until the highest office is reached -- national in some cases, regional in others. Note that this is a lengthy process since a single straggler can cause significant delays. In today's interconnected world, however, people want much faster results, which are transmitted electronically (cf. Section~\ref{subsec:cyberpapervoting}).

\myparagraph{Accumulation.}
Once all the reports have reached the final office, the signatures are verified, and the results are accumulated. After that, the results are published, but they are not official yet; some time is typically allowed for objections and requests for recounts before the decision is final.

\subsection{Elections vs. Referendums}
Most countries support two kinds of popular votes: elections to select representatives and referendums to decide on critical issues. Elections usually have multiple candidates to choose and sometimes also allow write-ins, to vote for a person that does not appear on the ballot. A referendum, on the other hand, usually only allows two answers: yes and no. There are some countries where additional choices are also available, but mostly referendums require clear binary decisions for critical issues (e.g., Brexit~\cite{brexitlaw}). 

Referendums and elections can be held for all levels of the hierarchy shown in Figure~\ref{fig:hirarchy}, e.g., for the election of the city council, the national parliament, or a referendum on joining the European Union. Depending on the hierarchy level, more votes must be transmitted and accumulated over more levels, leading to larger delays before the result is final. 
As representatives are usually elected for and within a district, there is no need for the federal office to accumulate all counts, while for a referendum it is -- indicated by the dashed lines in \Cref{fig:hirarchy}. 
An exception is, for example, the nation-wide elected president of France, who is elected by popular majority across all eligible voters in France.

\subsection{Cyber in Paper Voting}
\label{subsec:cyberpapervoting}
Paper voting promises purely paper-based procedures. However, recent demand for faster notification has led to several electronic component-replacements. In most reviewed countries, the results from voting stations are transmitted not only via postal mail but also using electronic communication channels (e.g., E-Mail, telephone, or specialized software). The results from such an electronic channel are not regarded as final; nevertheless, they are forwarded and accumulated to form \emph{official preliminary results}. Compared to final results, counting officials do not await all results of their subordinates before sending them on to the next office. After only basic feasibility checks, preliminary results are immediately forwarded and published by the highest offices (e.g., by the voting commission). News agencies use this information just hours after the polls close.

Exit polls provide even earlier projections: voters are asked for their decision after they have cast their vote. Exit polls cannot provide accurate estimations of the outcome since voters can refuse to participate in the survey leading to statistical uncertainty. However, the results of exit polls can be used for forecasts before the polls close. Similarly to preliminary results, exit polling data is usually transmitted electronically to a central entity that accumulates the results and publishes projections. Exit polls are typically conducted by private companies, in contrast to the official preliminary results.

Unlike exit polls, which tend to be unreliable, preliminary (complete) results are typically accepted and perceived as final by the public. Adversarial modification of the preliminary results can therefore have a significant impact. 

% Abstimmung vs. Election
% - Abstimmungen are more eligible to attack.

% - different modi
% - Four different kinds of results:
%   - Predictions, they suck 
%   - Exit Polls
%   - Preliminary results
%   - Final results

\subsection{Regulations on Accumulation of Preliminary Results}
%We examined the legal requirements for the transmission of (preliminary) results of elections in multiple nations. 
The regulations on the transmission of (preliminary) results vary widely across the reviewed countries. In Germany, Switzerland, and Austria, the election laws stipulate how preliminary results must be collected and transmitted as fast as possible~\cite{germanyvotinglaw,PRA}. However, they do not require any integrity or authenticity guarantee for the transmitted report. France, Estonia, and the United Kingdom do not define preliminary results in their laws at all. There, the local results must be publicly proclaimed in the voting station after the count is finished~\cite{francecodeelectoral,ukvotinglaw,estoniavotinglaw}, where private companies collect the results and provide the public with preliminary results and projections. In the United States, the individual states deploy different mechanisms; some specify the transmission of preliminary results while others rely on private companies. In general, no election law (that we studied) requires any guarantee regarding the integrity or authenticity of electronically transmitted results. Recently, some agencies have started publishing guidelines for electronic communication in paper voting systems, e.g., in the United States of America~\cite{us2015voluntary}. However, our inquiries to various election offices indicate that said recommendations are far from being implemented.

\section{Attacks on Preliminary Results}\label{sec:attack_vectors}

In this section, we investigate the regulations on the transmission of \textit{preliminary} results in paper voting and highlight various attack vectors on the insecure communication channels. We stress that once the paper trail arrives, the fraudulent preliminary results will be corrected. Nevertheless, preliminary results can exhibit a lot of influence on the world. 
While many countries have strong regulations on the paper-trail of an election to prevent tampering with the final result, preliminary results are often not as strictly governed. However, preliminary results are becoming increasingly significant, and they lead to extensive political and economic consequences. We stress that the problem of authenticity  has been mitigated for many years with secure channels~\cite{diffie1976new,merkle1978secure}, but in practice, many countries are still vulnerable to attacks on their digital communication of election results. 

\label{sec:pcwahl_scandal}

Recent reports in Germany~\cite{pcwahl} and the Netherlands~\cite{dutchscandal} highlight the vulnerable systems in place today. In the Netherlands, all results were transmitted entirely electronically without a paper trail unless someone objects the result. In Germany, only the preliminary results could be manipulated since the paper trail is always sent via paper mail. However, the Chaos Computer Club demonstrated how preliminary results for roughly half of the electors of Germany could have been manipulated due to grossly negligent software practices~\cite{pcwahl}. 

% Germany is a similar mess like das Helvetische Kleinreich. 
% \red{International comparison, how effected are the countries we have looked into?}
% \red{PC-Wahl}

\subsection{Attack Vectors}
\label{sec:new_attacks}
% We consider counting officials that count perfectly and transmit the correct preliminary and final results. Bribes and extortion aimed at counting officials are out of scope. The final result \emph{based on the paper trail} is always assumed to be correct. 
In this work, we only investigate attacks on preliminary results. While final results may be influenced by, e.g., bribing officials, we do not consider such attacks. We assume that election officials count and transmit votes correctly. Therefore, the final result based on the paper-trail is trusted.
Exit polls and preliminary results, however, are transmitted over electronic channels (e.g., VoIP or email) that can be manipulated. There are various types of communication channels used around the world, and we will discuss several attack vectors that might or might not apply to the respectively chosen channel. In certain cases, only a single municipality that still relies on insecure communication, if altered, can change close preliminary results of a national election or referendum (cf. \Cref{sec:impact-preliminary-results}).
Note that these manipulations must be performed blindly, i.e., an adversary does not know how close an election or a referendum will be or even what the final outcome will be. Nevertheless, blind attacks could still lead to manipulated preliminary results. 

\myparagraph{Tampering with Reports.}
Modifying the in-flight preliminary results is the most naive attack vector. Converting a few percents of votes in the report of a single municipality can easily lead to a swing in the national preliminary result. For example, swapping the yes and no vote-counts in the preliminary report will lead to an incorrect total preliminary result. The modification will be discovered as soon as the paper trail is processed (or an official notices the mistakes), but until then, the public will accept the fraudulent result.
The adoption of an authenticated and integrity protected communication channel would prevent any modification. However, our case study in Section~\ref{sec:casestudy-switzerland} shows that many constituencies use traditional communication means, such as telephone and email, and do not employ appropriate security mechanisms.

\myparagraph{Delaying Reports.}
Any attacker that controls any critical part of the communication channel (e.g., mail server, internet router, etc.) can delay any transmission of a preliminary report. The delaying attack becomes even stronger if the adversary is able to learn the delayed results and abuses this knowledge for individual profit. Even without knowing the result, one can delay results from a constituency where there is a strong expectation for a specific outcome. As an example, an adversary learns from polls that cities are strongly against a specific proposal while rural areas support it. Delaying the results from cities change the preliminary totals to dip temporarily in a chosen direction. Analysts will potentially warn that the results from cities did not arrive yet, but the public perception might already be affected.
We stress that the delaying attack works for various communication channels, and even secure channels are vulnerable.

\myparagraph{Front-Running with Fake Reports.}
While a specific voting office is still counting, a real-looking but maliciously forged preliminary report could be sent to the accumulating office before the actual report.
If it is accepted as valid, the central accumulating office will take the fraudulent preliminary results and forward them to the next higher instance. When the real preliminary results arrive, the error will most likely be detected and corrected, but the damage is already inflicted. The time during which the fraudulent results are considered as correct might be minimal, but the attacker knows that these results are fake, and can, therefore, adjust his strategy to profit from the incident (cf. Section~\ref{sec:impact-preliminary-results}). Note that the communication channel must allow arbitrarily forged messages for the front-running attack to work. 
%However, even a fully authenticated channel might not solve the problem as an official-looking E-Mail might trick an accumulating official to include results not sent over the correct channel.

% As an example, an adversary can create a forged report and send it to the supervising office before the official count has been finished. The time gap where the fraudulent result is used until the real count will be received can be exploited for personal gain (cf. \Cref{sec:impact-preliminary-results}).

\subsection{Feasibility of Attacks}
The feasibility of attacks depends significantly on the infrastructure and protocols in use. Nonetheless, a single, not explicitly involved person is usually in complete control of the communication infrastructure, and could potentially tamper with the accumulation process. If, for example, the preliminary results are transmitted via common email, an operator of any email server involved can drop and forge messages arbitrarily. The same is valid for telephone or Internet providers. 
% Phone numbers can be spoofed\cite{d} and voices can be artificially imitated\cite{d}.
A simple denial of service attack can facilitate a delaying attack: Delaying the reports by shutting down the communication channels for a few minutes might be enough to leverage financial gain. Front-running does not even require the ability to tamper with the communication channels. Knowledge about the format of vote-count reports and simple spoofing techniques are sufficient to preempt real reports. 

Many more sophisticated solutions suffer similar weaknesses~\cite{clarke2017e2eInsufficient}. WhatsApp~\cite{whatsappwhitepaper}, for example, is end-to-end encrypted, meaning its servers cannot read user-exchanged messages.
In the case of a man-in-the-middle impersonation attack, users only see an innocent ``security code has changed'' message~\cite{whatsappdeniesbackdoor,whatsappbackdoor}. However, the same message is also displayed for legitimate reasons, e.g., when a user re-installs WhatsApp or buys a new phone.
% However, regeneration of encryption-keys happens often, e.g., when a user re-installs WhatsApp, and affected users receive only an innocent "security code has changed" message~\cite{whatsappdeniesbackdoor,whatsappbackdoor}. 
To rule out an impersonation attack and reestablish confidentiality, electoral officials would once more need to compare their encryption-keys via an out-of-band channel. Such an impersonation attack could, for example, be run by a sufficiently privileged WhatsApp employee. 

% which is hard to interpret for normal users. Besides that this is prone to human error, these security messages need to be manually activated as well. 
%temporary encryption keys still lay on the servers to simplify offline-messaging\cite{d}. Therefore, a sufficiently privileged employee could forge and drop arbitrary messages between the users. That renders all three before-mentioned attacks possible. 

Avoiding such attacks is non-trivial. While tampering with multiple communication channels parallel increases the difficulty, it usually boils down to run the before-mentioned attacks simultaneously. 
%The fact that such processes are prone to human error\cite{Sheng:2010:phishing} might even let small discrepancies, e.g., a strange sender-address in an email, to slip through. 
Moreover, next to a technical solution, electoral officials also require specialized training to detect and handle a potentially malicious actor. 

% The presence of human errors\cite{Sheng:2010:phishing} and potential malicious actors requires specialized training for electoral officials

%Tampering with multiple communication channels in parallel is more difficult, but it usually boils down to run the before-mentioned attacks simultaneously. The fact that such processes are prone to human error\cite{Sheng:2010:phishing} might even let small discrepancies, e.g., a strange sender-address in an email, to slip through. Avoiding such attacks is non-trivial and next to a technical solution, the officials require specialized training to recognize suspicious events as well.
%How about:
%Avoiding such attacks is non-trivial since next to a technical solution, the human operator plays an important role. Officials must be trained to detect potential attacks and follow up on irregularities.

\section{Case Study: Switzerland}
\label{sec:casestudy-switzerland}

% * (no title)
% ** Short intro into political structure of CH
% ** election and referenda
% ** popular majority, cantonal majority
% * Transmission Systems
% ** Table 1 (or table 1a)
% ** Methodology for table 1
% * Historical Discrepancies
% ** Table 2 (or table 1b)
% ** Methodology for table 2
% * Potential Attacks in the Past
% ** on Popular Majority (bold paragraph begin) 
% ** on Cantonal Majority
% ** Little risk we conjecture, young padawan
 
We have chosen Switzerland for our case study, a direct democracy where the people decide on policies directly with 5.3M citizens eligible to vote~\cite{swiss-citizens}, as it has a flexible election and referendum mechanism with around ten distinct referendums per year. 
% \footnote{Out of 8.1M inhabitants, 23.7\% do not have Swiss citizenship~\cite{swiss-citizens}, 15.3\% have it but are under-age or crazy, plus 0.75M living abroad but still eligible to vote.}.
The constitution can only be changed if the majority of the eligible voters accept the change in a nation-wide referendum, called \emph{people's initiative}. Similar mechanisms are in place to object laws passed by the two legislative chambers of Switzerland. In this study, both cases are subsumed under the term \textit{referendum}.  

Politically, the Swiss federation consists out of 26 states, so-called \textit{cantons}. These are politically autonomous, and the organization of elections and referendums lies within their sovereignty. Each canton has its own way of conducting and transmitting voting results, and there is no significant effort to unify these systems. 
% This is what Swiss citizens call \textit{Kantönligeist} (\textit{cantonal spirit}, negative idiom describing insular and provincial thinking). 
\Cref{fig:switzerlandcomparison}a summarizes the cantons of Switzerland, including the number of eligible voters.
% Cantons are abbreviated using a two-letter identifier (e.g., BE for Bern). For a summary of all cantons including the identifier and number of eligible voters, see \Cref{sec:canton_summary}.

In federal elections, voters elect representatives for every canton individually. As there is no need for an accumulation of votes on a federal level, the impact of manipulated preliminary results of elections is limited to the canton's representatives.  
%For federal elections, the accumulation of votes stops at the state level (cantons) as even the candidates for a federal position are provided by a specific canton. Therefore, the impact of manipulated preliminary results of elections is limited to the canton's representatives. 
Nation-wide referendums, however, are aggregated on the federal level, and a tiny change in the outcome of one canton can flip the result of close referendums completely.
Any change in the federal constitution requires two majorities to pass: the popular majority of all votes of every participating citizen, and the majority of the cantons. The latter is composed of a vote per canton, 23 votes in total\footnote{Due to historic reasons, there are 6 cantons with only half a vote, casting together 3 of these 23 votes.}, where each vote is cast by the popular majority in each canton.
% where the majority of each canton casts a single vote, 23 votes in total\footnote{}. 
A referendum may win the popular majority due to an advantage in the highly populated cantons but fail to reach cantonal majority as many smaller cantons reject the referendum. With country-side cantons usually having fewer inhabitants, this leads to a balance between the densely populated urban cantons and more sparsely populated rural areas.

%\red{Even in case that such incidents happen, Switzerland is a stable democracy most likely able to handle them.}

% \begin{table*}
%     \centering
%     % \resizebox{\columnwidth}{!}{\input{tables/version3.tex}}
%     \input{tables/version3.tex}
% \end{table*}

% \begin{SCfigure*}
%   \centering
%   %\caption{ ... caption text ... }
%   \input{tables/version3.tex}
%           \caption{
%         % \begin{description}
%         %     \item[\Letter] Email.
%         %     \item[\Telefon] Telephone.
%         %     \item[\Faxmachine] Fax.
%         % \end{description}
%           The voting transmission mechanisms in all cantons of Switzerland for preliminary results. 'And' and 'Or' mean both or one of the transmission mechanisms must be used.}
%         \label{fig:switzerlandcomparison}
% \end{SCfigure*}

\subsection{Transmission Systems}

The final results of the vote are obtained by written paper-protocols that are signed by multiple members of each voting station, usually members of several political parties, and then sent by postal services to the next instance. The first accumulation of the paper report results usually takes 3-5 days, at most 13 days, followed by an objection period of 3 days~\cite{BGPR}. 
Simultaneous to these final results, there is an official accumulation of preliminary results, which are usually published on the day the vote takes place. 
%As shown in \Cref{sec:marketsreacttoprelimresults}, far-reaching decisions are already taken immediately after the preliminary results are published. 

On a federal level, current laws and regulations only specify that preliminary results must be published and that the transmission method must be either fax, phone, or teleprinter~\cite{PRA,PRR1}\footnote{Since the first publication of this work, this law has been updated to \textit{transmit preliminary results in electronic form} (valid in law since July 2019).}. As the law dates back to 1997, it does not consider the Internet appropriately; nevertheless, Internet-based communication methods (e.g., email) are used in practice as well. 
In accordance, the cantonal regulations mostly follow the federal ones except for some small adjustments: For example, the cantons of Appenzell-Innerrhoden and Glarus require the transmission of the results over an additional channel~\cite{AI,GL}, and the canton of Graubünden limits the communication to phone only~\cite{GR}. 

In practice, every canton provides its own solution to transmit and collect the preliminary vote-counts after the counting is finished. The involved technologies were inquired from each of the chancelleries individually by mail, phone, or personal interview, and in a few cases online-resources~\cite{regharm,sedex}, and are presented in \Cref{fig:switzerlandcomparison}b. In essence, the preliminary results are often transmitted using several known-to-be-weak transmission methods like email, telephone, and fax, which in their common form do not provide message authentication or integrity mechanisms. 
Many cantons apply individual dedicated software for voting results transmission, which have not passed a public security review. Given that 8 different dedicated software solutions are in use throughout Switzerland, the potential attack surface is rather large.
%
%The information for the federal level, namely its transmission method \textit{sedex}, was inquired from the federal chancellery, the federal office for statistics, and publications online\cite{regharm,sedex}.
Following the preliminary publication of our report, the chancellery of canton St. Gallen conducted their own survey on cantonal transmission systems, confirming our results~\cite{umfragesse}.

\begin{table*}[tbp]
\newcommand{\dainput}{
    \resizebox{1.01\textwidth}{!}{
        \hspace{-9.5pt}

\newcommand{\bcmwidth}{0.75pt}%  % \bigcmidrulewidth
\newcommand{\mypartsep}{0.5em}%  % \bigcmidrulewidth
\newcommand{\mcc}[1]{\multicolumn{1}{c}{#1}}% % short for multicolumn-centered
\newcommand{\mcd}[1]{#1 & #1}%
\newcommand{\sitrox}{Sitrox\cite{Sitrox}}%
\newcommand{\bewas}{Bewas\cite{Bedag}}%
\newcommand{\sygev}{SyGEV\cite{SyGEV}}%
\newcommand{\sesam}{Sesam\cite{Sesam}}%
\newcommand{\wabsti}{Wabsti\cite{abraxas}}%
\newcommand{\votel}{Votel\cite{votel}}%
\newcommand{\votelec}{Votelec\cite{votelec}}%
\newcommand{\adminvotel}{AdminVotel\cite{adminvotel}}%
\newcommand{\sedex}{Sedex\cite{sedex}}%
\newcommand{\tsep}{\,}
\newcommand{\taband}{\texttt{and}}%
\newcommand{\tabor}{\texttt{or}}%
\begin{tabular}{@{}lrlcclrrr@{}}%
\cmidrule[\bcmwidth]{1-2} \cmidrule[\bcmwidth]{4-5} \cmidrule[\bcmwidth]{7-9} % \toprule
    \multirow{2}{1em}[-4pt]{Canton} & \hspace{-6em}\multirow{2}{6.5em}[-4pt]{Eligible Voters\\\hfill in 2019~\cite{swiss-citizens}} &\phantom{\hspace{\mypartsep}}& \multicolumn{2}{c}{Transmission}  &\phantom{\hspace{\mypartsep}}& \multicolumn{3}{c}{Largest Discrepancy}   \\ \cmidrule{4-5} \cmidrule{7-9}
                       &     && Election  & Referendum  && \multicolumn{1}{c}{Votes} & \multicolumn{1}{c}{~~Total} & Percent     \\ \cmidrule[\bcmwidth]{1-2} \cmidrule[\bcmwidth]{4-5} \cmidrule[\bcmwidth]{7-9} 
    Aargau                 & 414\tsep{}745   &&  \mcd{\sitrox}                         && 1\tsep{}090   & 131\tsep{}179  & 0.83\% \\ %& 1,404         \\
    Appenzell Ausserrhoden\hspace{-0.8em} & 38\tsep{}498    &&  \mcd{\Letter ~\taband{} \Faxmachine}  && 0      & 16\tsep{}466   & 0.00\% \\ %& 243           \\
    Appenzell Innerrhoden\hspace{-0.8em}  & 11\tsep{}565    &&  \mcd{\Letter ~\tabor{} \Telefon}      && 0      & 3\tsep{}800    & 0.00\% \\ %& 173           \\
    Basel-Landschaft       & 187\tsep{}863   &&  \Letter  & \Telefon                   && 206    & 81\tsep{}622   & 0.25\% \\ %& 518           \\
    Basel-Stadt            & 113\tsep{}717   &&  \mcd{\Letter}                         && 2      & 53\tsep{}918   & 0.00\% \\ %& 37            \\
    Bern                   & 729\tsep{}203   &&  \mcd{\bewas}                          && 1\tsep{}605   & 316\tsep{}539  & 0.51\% \\ %& 5,959         \\
    Fribourg               & 196\tsep{}027   &&  \mcd{\sygev}                          && 1\tsep{}230   & 82\tsep{}176   & 1.50\% \\ %& 1,671         \\
    Genève                   & 248\tsep{}915   &&  \mcd{\Telefon}                        && 142    & 108\tsep{}651  & 0.13\% \\ %& 282           \\
    Glarus                 & 26\tsep{}268    &&  \sesam & \Letter ~\taband{} \Telefon  && 19     & 9\tsep{}040    & 0.21\% \\ %& 685           \\
    Graubünden             & 137\tsep{}126   &&  \mcd{\sesam}                          && 690    & 58\tsep{}784   & 1.17\% \\ %& 7,105         \\
    Jura                   & 51\tsep{}936    &&  \mcd{\Telefon}                        && 548    & 20\tsep{}178   & \textbf{2.72\%} \\ %& 838           \\
    Luzern                 & 271\tsep{}143   &&  \Telefon & \sesam                     && 276    & 112\tsep{}269  & 0.25\% \\ %& 1,493         \\
    Neuchâtel              & 111\tsep{}304   &&  \mcd{\sygev}                          && 891    & 46\tsep{}970   & 1.90\% \\ %& 803           \\
    Nidwalden              & 30\tsep{}810    &&  \mcd{\sesam}                          && 100    & 10\tsep{}915   & 0.92\% \\ %& 276           \\
    Obwalden               & 26\tsep{}244    &&  \mcd{\sesam}                          && 0      & 9\tsep{}661    & 0.00\% \\ %& 491           \\
    Schaffhausen           & 51\tsep{}036    &&  \mcd{\Letter ~\tabor{} \Telefon ~\tabor{} \Faxmachine} && 50     & 26\tsep{}625   & 0.19\% \\%& 298 \\
    Schwyz                 & 102\tsep{}145   &&  \mcd{\wabsti}                         && 0      & 37\tsep{}313   & 0.00\% \\ %& 908           \\
    Solothurn              & 177\tsep{}292   &&  \mcd{\wabsti}                         && 1\tsep{}076   & 92\tsep{}400   & 1.16\% \\ %& 791           \\
    St. Gallen             & 317\tsep{}969   &&  \mcd{\wabsti}                         && 1\tsep{}012   & 162\tsep{}996  & 0.62\% \\ %& 2,026         \\
    Thurgau                & 206\tsep{}118   &&  \mcd{\wabsti}                         && 570    & 71\tsep{}626   & 0.80\% \\ %& 991           \\
    Ticino                 & 218\tsep{}580   &&  \mcd{\votel}                          && 886    & 84\tsep{}892   & 1.04\% \\ %& 2,812         \\
    Uri                    & 31\tsep{}928    &&  \mcd{\Telefon ~\taband{} \sesam}      && 104    & 9\tsep{}583    & 1.09\% \\ %& 1,077         \\
    Valais                 & 216\tsep{}041   &&  \mcd{\adminvotel}                     && 1\tsep{}232   & 118\tsep{}091  & 1.04\% \\ %& 5,224         \\
    Vaud                  & 428\tsep{}569   &&  \mcd{\votelec}                        && \textbf{3\tsep{}974}   & 177\tsep{}616  & \textbf{2.24\%} \\ %& 3,212         \\
    Zug                    & 74\tsep{}803    &&  \wabsti & \Letter                     && 0      & 34\tsep{}444   & 0.00\% \\ %& 239           \\
    Zürich                 & 907\tsep{}623   &&  \mcd{\wabsti}                         && 1\tsep{}400   & 498\tsep{}786  & 0.28\% \\ \cmidrule{1-2} \cmidrule{4-5} \cmidrule{7-9} %& 1,729         \\ \midrule
    Switzerland            & 5\tsep{}283\tsep{}556 &&  \mcd{\sedex}                          && 3\tsep{}843   & 2\tsep{}039\tsep{}548 & 0.19\% \\ %& 41,285        \\
    \cmidrule[\bcmwidth]{1-2} \cmidrule[\bcmwidth]{4-5} \cmidrule[\bcmwidth]{7-9}
\end{tabular}

        %\input{tables/version5abrv.tex}
        % \newline
    }
}
\newcommand{\middleseperator}{1em}
  \scalebox{0.95}{
\newcommand{\bcmwidth}{0.75pt}%  % \bigcmidrulewidth
\newcommand{\mypartsep}{1em}%  % \bigcmidrulewidth
\newcommand{\mcc}[1]{\multicolumn{1}{c}{#1}}% % short for multicolumn-centered
\newcommand{\mcd}[1]{#1 & #1}%
\newcommand{\sitrox}{Sitrox\cite{Sitrox}}%
\newcommand{\bewas}{Bewas\cite{Bedag}}%
\newcommand{\sygev}{SyGEV\cite{SyGEV}}%
\newcommand{\sesam}{Sesam\cite{Sesam}}%
\newcommand{\wabsti}{Wabsti\cite{abraxas}}%
\newcommand{\votel}{Votel\cite{votel}}%
\newcommand{\votelec}{Votelec\cite{votelec}}%
\newcommand{\adminvotel}{AdminVotel\cite{adminvotel}}%
\newcommand{\sedex}{Sedex\cite{sedex}}%
\newcommand{\taband}{\texttt{and}}%
\newcommand{\tabor}{\texttt{or}}%
\begin{tabular}{lrlcclrrr}%
\cmidrule[\bcmwidth]{1-2} \cmidrule[\bcmwidth]{4-5} \cmidrule[\bcmwidth]{7-9} % \toprule
    \multirow{2}{*}[-4pt]{Canton} & \hspace{-0.8cm}\multirow{2}{6.5em}[-4pt]{Eligible Voters\\\hfill in 2019~\cite{swiss-citizens}} &\phantom{\hspace{\mypartsep}}& \multicolumn{2}{c}{Transmission}  &\phantom{\hspace{\mypartsep}}& \multicolumn{3}{c}{Largest Discrepancy [votes]}   \\ \cmidrule{4-5} \cmidrule{7-9}
                       &     && Election  & Referendum  && Incorrect & Total & Percent     \\ \cmidrule[\bcmwidth]{1-2} \cmidrule[\bcmwidth]{4-5} \cmidrule[\bcmwidth]{7-9} 
    Aargau                 & 414,745   &&  \mcd{\sitrox}                         && 1090   & 131179  & 0.83\% \\ %& 1,404         \\
    Appenzell Ausserrhoden & 38,498    &&  \mcd{\Letter ~\taband{} \Faxmachine}  && 0      & 16466   & 0.00\% \\ %& 243           \\
    Appenzell Innerrhoden  & 11,565    &&  \mcd{\Letter ~\tabor{} \Telefon}      && 0      & 3800    & 0.00\% \\ %& 173           \\
    Basel-Landschaft       & 187,863   &&  \Letter  & \Telefon                   && 206    & 81622   & 0.25\% \\ %& 518           \\
    Basel-Stadt            & 113,717   &&  \mcd{\Letter}                         && 2      & 53918   & 0.00\% \\ %& 37            \\
    Bern                   & 729,203   &&  \mcd{\bewas}                          && 1605   & 316539  & 0.51\% \\ %& 5,959         \\
    Freiburg               & 196,027   &&  \mcd{\sygev}                          && 1230   & 82176   & 1.50\% \\ %& 1,671         \\
    Genf                   & 248,915   &&  \mcd{\Telefon}                        && 142    & 108651  & 0.13\% \\ %& 282           \\
    Glarus                 & 26,268    &&  \sesam & \Letter ~\taband{} \Telefon  && 19     & 9040    & 0.21\% \\ %& 685           \\
    Graubünden             & 137,126   &&  \mcd{\sesam}                          && 690    & 58784   & 1.17\% \\ %& 7,105         \\
    Jura                   & 51,936    &&  \mcd{\Telefon}                        && 548    & 20178   & \textbf{2.72\%} \\ %& 838           \\
    Luzern                 & 271,143   &&  \Telefon & \sesam                     && 276    & 112269  & 0.25\% \\ %& 1,493         \\
    Neuenburg              & 111,304   &&  \mcd{\sygev}                          && 891    & 46970   & 1.90\% \\ %& 803           \\
    Nidwalden              & 30,810    &&  \mcd{\sesam}                          && 100    & 10915   & 0.92\% \\ %& 276           \\
    Obwalden               & 26,244    &&  \mcd{\sesam}                          && 0      & 9661    & 0.00\% \\ %& 491           \\
    Schaffhausen           & 51,036    &&  \mcd{\Letter ~\tabor{} \Telefon ~\tabor{} \Faxmachine} && 50     & 26625   & 0.19\% \\%& 298 \\
    Schwyz                 & 102,145   &&  \mcd{\wabsti}                         && 0      & 37313   & 0.00\% \\ %& 908           \\
    Solothurn              & 177,292   &&  \mcd{\wabsti}                         && 1076   & 92400   & 1.16\% \\ %& 791           \\
    St. Gallen             & 317,969   &&  \mcd{\wabsti}                         && 1012   & 162996  & 0.62\% \\ %& 2,026         \\
    Tessin                 & 218,580   &&  \mcd{\votel}                          && 886    & 84892   & 1.04\% \\ %& 2,812         \\
    Thurgau                & 206,118   &&  \mcd{\wabsti}                         && 570    & 71626   & 0.80\% \\ %& 991           \\
    Uri                    & 31,928    &&  \mcd{\Telefon ~\taband{} \sesam}      && 104    & 9583    & 1.09\% \\ %& 1,077         \\
    Waadt                  & 428,569   &&  \mcd{\votelec}                        && \textbf{3974}   & 177616  & \textbf{2.24\%} \\ %& 3,212         \\
    Wallis                 & 216,041   &&  \mcd{\adminvotel}                     && 1232   & 118091  & 1.04\% \\ %& 5,224         \\
    Zug                    & 74,803    &&  \wabsti & \Letter                     && 0      & 34444   & 0.00\% \\ %& 239           \\
    Zürich                 & 907,623   &&  \mcd{\wabsti}                         && 1400   & 498786  & 0.28\% \\ \cmidrule{1-2} \cmidrule{4-5} \cmidrule{7-9} %& 1,729         \\ \midrule
    Switzerland            & 5,283,556 &&  \mcd{\sedex}                          && 3843   & 2039548 & 0.19\% \\ %& 41,285        \\
    \cmidrule[\bcmwidth]{1-2} \cmidrule[\bcmwidth]{4-5} \cmidrule[\bcmwidth]{7-9}
\end{tabular}

}
  
\small 
\noindent\hspace{8em}(a) \hspace{19.5em} (b) \hspace{18.5em} (c)
\normalsize
    
  \vspace{1pt}
  \caption{
        \textbf{(a)} The different cantons of Switzerland with the number of eligible voters in 2019, extended by the total of Switzerland in the last row.
        \textbf{(b)} The voting transmission mechanisms for preliminary results from municipalities to cantons in Switzerland as of January 2019. The last row indicates the transmission mechanism which is used from the cantonal chancelleries to the federal chancellery. 
        % Several cantons like Genf or Jura only rely on telephone for transmission. 
        Note that there are 11 different transmission systems leading to a large attack surface.
        \textbf{(c)} The largest discrepancy of counted votes between the preliminary and final results compared to the total number of votes per canton for referendums held from Sept. 2009 to May 2019, 84 referendums in total. Note that the maximal discrepancies of each canton might have occurred in different referendums. 
        \textbf{Legend:}  Fax (\Faxmachine),  Email (\Letter), Telephone (\Telefon),  either one (\texttt{or}), or both simultaneously (\texttt{and}).
        % \vspace{-3em}
    }  \label{fig:switzerlandcomparison}
\end{table*}

\subsection{Historic Discrepancies}
We analyzed historical discrepancies in all 84 referendums of Switzerland in the last ten years, from \nth{27} Sept. 2009 until \nth{19} May 2019 (from referendum No. 543 to No. 626). We collected preliminary results from newspaper articles from the next day,
%\footnote{We used the newspapers Tagesanzeiger and Neue Z\"urcher Zeitung which in turn get their data from Keystone-SDA.}
and the final results from the official report by the federal office of statistics. Remarkably, not a single preliminary result was without deviations. However, we believe all of these inconsistencies originate from human errors and not from attacks.

The maximum relative deviation of each canton for a single referendum is shown in \Cref{fig:switzerlandcomparison}c. Note that the entries in \Cref{fig:switzerlandcomparison}c are not correlated: the referendum with the maximum discrepancy for the canton Jura might be a different referendum than the one for Bern. Single cantons show a deviation of up to 2.72\%: The canton of Vaud erred once in 3974 counts. While the federal average discrepancy over the ten years is 0.03\%, the average over the maximum cantonal discrepancy per referendum is 0.43\%. These deviations are already larger than yes-no gaps of several close referendums. Additionally, we noted that discrepancies usually balance themselves, e.g., canton of Jura might make a mistake in one direction and canton of Bern in the other. However, if these mistakes coincide, there might even be a more significant discrepancy on the federal level.

% Geneva numbers = final-wrong array([85114, 24319]), final-true array([63980, 48010]) preliminary array([85103, 24312]), FORMAT: [yes, no]
In one case, the referendum from \nth{22} Nov. 2013 on the objection to the new law on epidemics, the final results in the canton of Geneva deviated by 41\% (44'821 of 109'433 totally cast votes) from the preliminary results. Eight months after, the final result was corrected, finally only differing by 18 votes from the initial preliminary results~\cite{genevaDesaster2011}. The overall outcome, however, did not change; neither in the canton of Geneva nor on the federal level.

\subsection{Examples of Vulnerable Referendums}
As illustrated previously, maliciously manipulating the result from a few municipalities by a few negligible percentage points is feasible. This section presents two historical examples of nation-wide referendums which were close enough to flip the outcome. We stress that the proposed attacks on these referendums are fictional and, to the best of our knowledge, did not occur.
Note that in these scenarios we consider an all-knowing attacker who already knows the outcome before the referendum. Therefore, the manipulations can be hand-tailored to change to final decision. In reality, the adversary would have to mount such attacks blindly and would potentially have to overshoot with the manipulations.
%For some referendums, the gap of yes and no votes lays within the normal fluctuations.
Although there are severe security risks in the transmission infrastructure, we do not consider the system in Switzerland to be at imminent risk of large manipulations. 
%For close referendums, however, attacks stay feasible.
For close referendums however, attacks stay feasible.

\myparagraph{Attack on Popular Majority.}
%This attack takes place during the vote "Volksentscheid gegen Asylmissbrauch" (referendum on abuse of the asylum system) from \nth{24} November 2002~\cite{asylmissbrauch}. It was rejected  by the popular majority with 1'119'342 to 1'123'550 votes. In contrast, the majority of the cantons of the vote was won. To flip the vote to acceptance, an attacker would need to change only 2104 votes.
%
%We change those votes in the canton Basel-Landschaft (BL) since it is the largest that uses e-mail to transmit preliminary voting results. Without an attacker's action, the referendum in BL was won with 50.3\%. In the case of manipulation, the vote would have been won as well, but now with 52.7\%. With these minor shifts, the outcome of the federal vote would have been flipped. 

This attack takes place during the vote on the ``Bundesgesetzes über Radio und Fernsehen (RTVG)'' (law on radio and television) from \nth{26} September 2014~\cite{bundesbeschluss2015radio}. It was accepted  by the popular majority with 1,128,522 to 1,124,873 votes. Since this referendum only proposed changes in the law but not the constitution, the majority of cantons was not required. To flip the vote to acceptance, an attacker would need to change only 1825 votes.
We change those votes in the canton Basel-Landschaft (BL) since it is relatively large and uses e-mail to transmit preliminary voting results. Without an attacker's action, the referendum in BL was rejected with 54.2\% \emph{no}-votes. In the case of manipulation, the vote would have been rejected as well, but now with 56.5\%. With these minor shifts, the outcome of the federal referendum would have been flipped.

\myparagraph{Attack on Cantonal Majority.}
The second attack aims to flip the cantonal majority. We target the referendum on family support from \nth{3} March 2013~\cite{bundesbeschluss2013familienpolitik}. The popular majority was achieved with a 54.3\% yes votes on the federal level. However, only 10 out of 23 cantons voted to accept the referendum. To swing the result, the decision of two cantons would be needed to flip.
We choose to attack the cantons Graub\"unden and Zug. In the canton of Graub\"unden the referendum was rejected with 36,130 against 37,920 and in Zug with 17,703 against 19,570. In total, an attacker would have needed to flip half of the difference in each canton, 895 in Graub\"unden and 687 in Zug, totaling to 1582 votes.

% The second attack aims to flip the majority of the cantons. We target the decision on joining the United Nations from \nth{3} March 2002~\cite{uno}. On the federal level, the decision was won with 54.6\% regarding the popular majority, and 12 cantonal votes opposed by 11 cantonal votes regarding the majority of the cantons. If the two majorities do not agree, the vote is rejected. So the attack only must change a single full cantonal vote.
% %
% We choose to attack the canton of Zug because it is the closest canton that uses email or other non-authenticated methods. Originally, the cantonal vote was won with 55.2\%. To flip the vote to a negative outcome, i.e. a result below 50\%, we need to change only 2'438 of a total of 44'708 votes.

\subsection{Methodology}
The various transmission methods used throughout Switzerland were inquired from each of the chancelleries individually by mail, phone, or personal interview, and in a few cases online-resources~\cite{regharm,sedex}. 
%The historic preliminary results were gathered from newspaper articles on the day after\footnote{We used the newspapers Tagesanzeiger and Neue Z\"urcher Zeitung which in turn get their data from Keystone-SDA.}, and the final results from the official reports by the federal office of statistics.
The historic preliminary results were gathered from newspaper articles on the day after, namely Tagesanzeiger and Neue Z\"urcher Zeitung which get their data from Keystone-SDA. The historic final results come from the official reports by the federal office of statistics.

\section{Impact of Preliminary Results}
\label{sec:impact-preliminary-results}

During our investigations, many state officials claimed that a modification attack on preliminary voting results does not matter --- arguing that only the official accumulation of the hand-signed paper reports count. However, we conjecture that the public perceives this differently:
%While most state-officials claim that a modification attack on preliminary voting results does not matter -- arguing that only the official accumulation of the hand-signed paper reports count -- the public perceives this differently. 
In many cases, the preliminary results are regarded as final, usually due to ignorance or out of experience that there was never any significant discrepancy.
While from a formal-law viewpoint negligible, there are extensive decisions taken based on these preliminary results -- for example, reacting stock markets or foreign currency exchange rates, induced long-term political decisions that are taken before the results are final, or high-frequency trading that reacts immediately on very recent information. 
In the case of tampered preliminary results, significant harm may already have occurred, even if the fraud will be detected one or two days later. Moreover, the confidence in the voting process might suffer, which can lead to lowered trust in the democracy in general. Financial decisions based on false knowledge may bring significant loss, and far-reaching decisions might require a revocation. We will highlight three examples where preliminary results correlated significantly with economic reactions and political events hours or days before the final results were published.

% \red{mention predictions and exit polls as well}

\subsection{Market Insecurity after Brexit}
  
Our first example is the decay of the exchange rate of British pounds to US-Dollar in the night of the \nth{23} to \nth{24} June 2016. On that day, the ballots were counted on the referendum whether the United Kingdom shall invoke article 50 of the Treaty of the European Union -- which would start a lengthy process towards UK's exit from the European Union~\cite{brexitlaw} -- commonly referred to as \textit{Brexit}. The referendum was finally accepted with a majority of 51.9\%~\cite{brexitresult}. 

\begin{figure}[tbp]
  \centering
     \iftwocolumn
        \includegraphics[width=1.0\linewidth, trim=10.5pt 0 0 0, ]{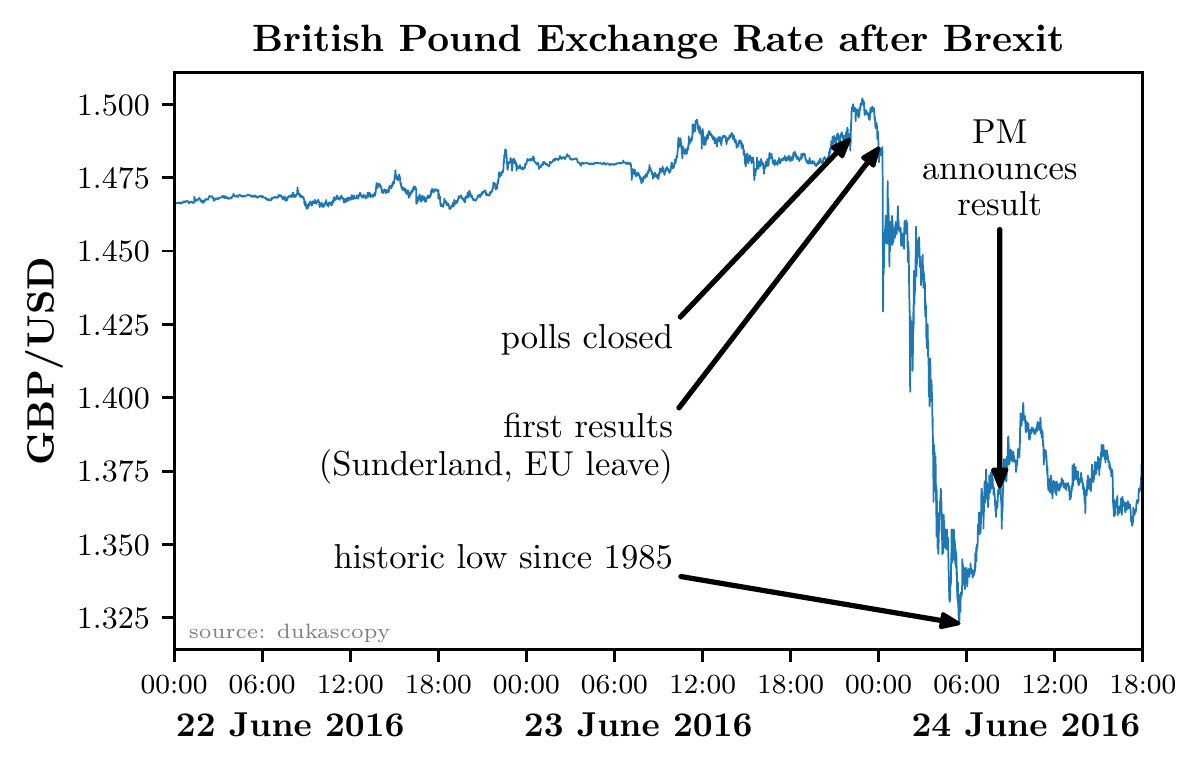}
     \else
        \includegraphics[width=0.75\linewidth, trim=10.5pt 0 0 0, ]{data/Brexit_23_06_2016/plot.pdf}
     \fi

    \vspace{-1.0em}

   \caption{The currency exchange rate of British Pounds vs. US-Dollar around the Brexit-referendum vote. (Source: dukascopy) %\spacesaver
    } \label{fig:brexitpounds}
    
\end{figure}

While the impact of the referendum on the British Pound was discussed the days before the Brexit referendum, in the night the votes were counted and the preliminary results were piece-wise published, the GBP/USD exchange rate fell to a historic low. \Cref{fig:brexitpounds} illustrates the development. The polls closed at 22:00 followed by a slight raise of the exchange rate. After the first election office (Sunderland) announced its result (EU leave), the exchange rate fell immediately, eventually reaching the lowest point since 1985. The next morning, 8:15am at \nth{24} June 2016, the prime minister of the United Kingdom spoke to the public announcing the final results\cite{cameronbrexit}. The market stabilized again, but the exchange rate was still over 8\% lower than the day before. 
% If the first result would have rejected the referendum (EU remain), the market might have reacted differently. 
% The market might have reacted differently, if the first result would have rejected the referendum (EU remain).
Another fist publication might have impacted the marktet differently. 

% old
% \begin{wrapfigure}[15]{r}{0.65\textwidth}
%   \centering\vspace{-0.5em}
%   %\vspace{-2.3em}
%      % trim={<left> <lower> <right> <upper>}
%      \includegraphics[width=0.66\textwidth, trim=10.5pt 0 0 0, ]{data/Brexit_23_06_2016/plot.pdf}

%   \caption{The currency exchange rate of British Pounds vs. US-Dollar around the Brexit-referendum vote.
%     } \label{fig:brexitpounds}
% \end{wrapfigure}\noindent

% \red{most likely PM statement the next morning is not a paper-trail, but electronically transmitted results}
% \red{imagine the different market behaviour if Sunderland would have falsely proclaimed "EU stay"}
\label{sec:marketsreacttoprelimresults}

\subsection{Federal Elections in Germany 2017}
\label{sec:german_election}
In Germany, there exist official preliminary results that are publicly announced usually the day the polls close. The final results are provided 10 to 30 days later by the \emph{Bundeswahlleiter} (federal election director). On the \nth{24} September 2017, the 2017 federal parliament election of Germany took place, a few weeks after the discovery of severe vulnerabilities in the vote transmission software (cf. \Cref{sec:pcwahl_scandal}). Preliminary results were available the same day, and the identical final result was approved 19 days later, on the \nth{12} October 2017.

According to the preliminary results, the social democratic party of Germany (SPD) lost a significant number of seats to the new right-wing party \textit{Alternative für Deutschland}. The SPD lost its previous advantage for the participation in the parliament's major governing alliance. Therefore, they announced party-internal changes and stated to go into opposition already on the \nth{24} September~\cite{germany1}, not a day after the preliminary results were published. The next day, insecurities in the stock markets followed due to the recent change in the country's power structure~\cite{germany2}. Four days after the preliminary results, the president of the United States of America, Donald J. Trump, congratulated the German Chancellor Angela Merkel for her reelection~\cite{germany3}. Finally, on the \nth{9} October 2017, still several days before the results were final, the newly formed government coalition fixed the date they planned to start the negotiations about their future coalition~\cite{germany4}.

While none of these actions were irreversible, they were all taken before the election results were final, and during the ongoing public discussion about insecure voting result transmission software (cf. \Cref{sec:pcwahl_scandal}). If, and this was not unlikely at that time~\cite{pcwahl}, the preliminary result would have been tampered with, such an error could have left a lasting negative impression among the people of Germany.

\begin{figure*}[tbp]
    \centering
    \vspace{-0.7em}
    
    \includegraphics[width=\textwidth]{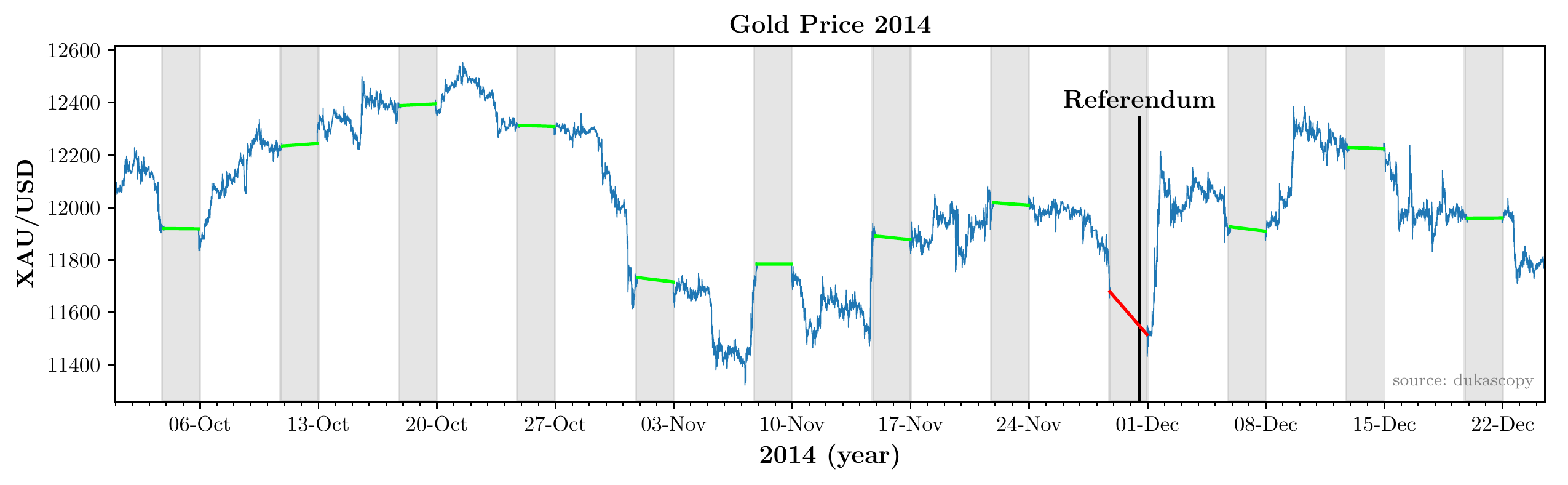}
    
    \vspace{-1.0em}
    \caption{The price of a troy ounce of gold in US-Dollar. The market is closed on weekends (gray underlays). The price has fallen significantly over the weekend where the gold referendum of Switzerland took place, indicated by the red line. The preliminary outcome was confirmed by the final result several days later.}
    % \spacesaver
    \label{fig:goldstandart}
\end{figure*}

\subsection{Vote on Gold Reserves in Switzerland}

In 2014, Switzerland held a referendum whether the Swiss National Bank (SNB) is obligated to store 20\% of its foreign-exchange reserves in gold. At that time, the SNB retained only 7.5\% of its foreign-exchange reserves in gold and, in case of acceptance of the referendum, would have been forced to buy around 1500 metric tons of Gold (worth around 60 Billion US-Dollar) within the next three years~\cite{schweiz1}. The referendum was rejected on \nth{30} November 2014 by a large majority; the final result followed several days later. 

This referendum became a major talking point internationally. Several News outlets discussed the referendum and showed its impact on the international gold price~\cite{golddiscussions}. \Cref{fig:goldstandart} pictures the exchange rate of a troy ounce of gold against US-Dollar. The exchange is closed on the weekends, hinted by the grey underlays in the figure. Usually, the exchange market opens on Mondays with roughly at the same price as it has closed the Friday before, indicated by the horizontal green lines.
In the weekend where the referendum took place, however, the markets opened with a significantly lower price, indicated by the red line. Experts attribute this decrease to the rejection of the referendum~\cite{schweiz2} whose preliminary result was published Sunday evening, days before the final one. In the case of adversarial tampering of the preliminary outcome, false investments would have followed, the gold market could have seen increased insecurity and instability, and the trust into the Swiss democracy might have suffered.

\section{Recommendations}
\label{sec:recommendations}
The primary issue in all attacks on preliminary voting reports is the missing authenticity and integrity of the reports transmitted; the broadly used email or telephone do not provide these properties in their common form. In this section, we show how simple already existing technology can be used to mitigate the previously mentioned attacks. First, all communication channels need to guarantee authenticity and integrity, and second, keys must be handled securely, e.g., with smartcards.

For the transmission of preliminary voting results, signing the report to allow for subsequent verification must be mandatory. Moreover, each preliminary report needs to be uniquely attributed to a specific election and, in the case of intermediate result publications, contain a monotonic counted number to avoid replay attacks. Existing systems such as PGP~\cite{garfinkel1995pgp} already satisfy these requirements and are widely adopted. 
% or TLS~\cite{tls}

Such technologies rely on asymmetric encryption to verify the origin of messages based on a public key. If every vote-counting and aggregation entity holds an individual private key to sign its voting-reports, then the validity of any report can be verified using the corresponding public certificate. However, the distribution of such trusted certificates is not trivial.
Public Key Infrastructure (PKI)~\cite{cooper2008internet} is a perfect fit for the hierarchical structures in a typical country: The federal office could provide the root certificate and sign the certificates of the subordinate districts, who would then, in turn, sign the certificates of their municipalities. As long as the report is signed by a key belonging to the correct certificate, its authenticity can be validated via the chain of certificates, eventually reaching the trusted root certificate of the federal office. Many countries have already rolled out a PKI to secure their internal communication~\cite{swiss-pki,germaniapki}.

%All such technologies rely on asymmetric encryption to verify the origin of messages based on a public key. The distribution scheme of such public keys should be chosen to accommodate the use-case: Public Key Infrastructure (PKI)~\cite{cooper2008internet} is a perfect fit for hierarchies of countries. The federal office could provide the root certificate and sign the certificates of the subordinate districts, who would then, in turn, sign the certificates of their counties. As long as the report is signed by a key belonging to the correct certificate, its authenticity can be validated via the chain of certificates. 

The private keys belonging to the certificates are critical, as they allow an adversary to forge signatures on fraudulent reports. Additionally, any lost or compromised key must be revoked as soon as possible. In order to protect the private keys from leaking, a country could make use of technologies such as smartcards~\cite{rankl2004smart}, which encapsulate a private key in a credit-card like physical device and thereby making key extraction difficult~\cite{kocher1999differential}. The adversary needs to steal the smartcard to sign fraudulent reports, and a missing smartcard is detected quickly. Moreover, manipulating an election or referendum on a large scale would require several cards.

% We stress that, regardless of the transmission mechanism used, delaying the report is always possible without availability guarantees (synchronous channels).

% \red{sing the certs for every election again?}

% \red{Refer to the official recommendations ,e.g., US?}

% -we cannot prevent wring investments before all the results are collected -> delay attacks? 

% - problem: need authenticity an integrity of massages. pki. like TLS or dnssec.

% -deeelay attack. 

% Solution: 
% - Blockchain! (just kidding, i meant SGX)
% - each voting office: individual Smartcards with individual pub-cert signed by governement.
% - The public needs to ignore exit-polls and other non-official predictions.
% - voting office signs every (optimally machine-readable) statement with key on smartcard. Then it does not matter how they are transmitted, as the only important property is integraity.  

% Assumptions: 
% - smart card private keys are not leaked.
% - governement private cert is not leaked.
% - people do not lose their smartcard all the time.

% Advantages:
% - smart card not easy so duplicate/fake
% - even when single smart cards leak, they can be replaced, their key pub key revoked.
% - large scale attacks very difficult.

% Signature of everything with Smartcard. 

% \section{Ethics and Regulations}
% Due to the severe impact on the stability of whole societies, we do not have approval of our ethics-board for testing our hypothesis in a real-world setup.

\section{Conclusion}
\label{sec:conclusion}
This work examines the non-negligible time-gap between preliminary and final results of paper-based voting mechanisms across several countries. In most countries, the final results of elections and referendums are determined based on written and hand-signed paper-reports of the local results of each individual voting office, which are sent (by postal services) to the next higher aggregation office up to the topmost level that publishes the final result. As this process usually takes several days, faster aggregation systems for preliminary results have emerged, which use insufficiently secured digital transmission systems for preliminary voting results. As a result, the preliminary outcome of elections and referendums can be modified by an adversary.
While such tampering is unlikely to succeed for one-sided referendums or elections, very close decisions can be flipped by a small distortion of a regional result. Our case-study of Switzerland highlights several potentially vulnerable referendums, and given the reports about the software vulnerabilities of aggregation systems in Germany and the Netherlands, we conjecture similar issues in many other countries as well. 
While tampering with preliminary results should be discovered at the latest when aggregating the final results, we have investigated the substantial impact of modified preliminary results with several examples. In the case of the marginal Brexit referendum -- 51.9\% voted for leaving the EU -- the value of the British Pound crashed after the publication of the first preliminary result. In contrast, the market might have reacted differently if the first result would have proclaimed ``EU stay.'' 

%We do not consider established  democracies like Switzerland or Germany to be at imminent risk, since these provide working mechanisms to cope with such incidents. But reliable counting and aggregation mechanism are essential to maintain confidence in such an important pillar of modern democracies like elections and referendums. As Joseph Stalin, a famous demagogue very effective in retention of power, once presumably said: "I consider it completely unimportant who in the party will vote, or how; but what is extraordinarily important is this -- who will count the votes, and how."

% It’s not the people who vote that count. It’s the people who count the votes.

% \vfill
% \pagebreak 
% \small
%\bibliographystyle{ieeetr}
\bibliographystyle{plain}
\bibliography{paper}
\par\leavevmode
\end{document}